# Factual and Legal Risks Regarding Wireless Computer Networks

Maximillian Dornseif, Kay H. Schumann, Christian Klein

*There has been much talk about the risks to radio networks. How great is the actual danger and does criminal law protect against assaults on radio networks?*

Maximillian Drnseif

is working on his doctorate at the professorship for criminal law at the University of Bonn on the topic "Legal Facts of Computer Crime"
E-Mail: md@hudora.de

Kay H. Schumann

is an articled clerk. He is working on his doctorate at the professorship for criminal law at the University of Bonn on the topic "The Relevance of Electronic Payment according to Criminal Law"
E-Mail: kayschumann@mac.com

Christian Klein

is studying computer science at the University of Bonn and works as a Junior Consultant for c0re GmbH, Bonn.
E-Mail: ck@c0re.23.nu

## Introduction

In the past few months there have been many articles on the risks of Wave-LANs, the wireless computer networks according to the IEEE 802.11 standard [1]. The terms *drive-by-hacking* and/or *war driving* are being used. However, so far methodical inquests have concentrated on cryptographic problems [2]. In this article we will explain the significance of these problems in the context of protocol mechanisms and make an evaluation of the penal code in respect thereof. In addition, we will present the results of an assessment of the actual situation -- for this purpose we investigated the extension and configuration of Wave-LANs in the Bonn area.

## 1 Problem

Wireless communication differs from cabled communication by the applied medium of transmission. Regulating the access of data over cable is a relatively simple matter: Those with no cable access have no admission to the transmitted data without a considerable amount of effort.

The transmission medium for wireless communication is radio waves, which are broadcast circularly from the transmitter. Thereby, the power of reception subsides quadratically in relation to the distance from the transmitting source. The broadcast radius is weaker in urban areas and an estimation of propagation is unpredictable.

Even when propagation is partly controlled by a skilful selection of a private "network" of antennas and transmitters, the area of coverage does not stop at building or property boundaries. In practice this often leads to the possibility of regularly receiving institutional wireless communication on public property.

This problem had been known before DECT Standards were introduced in connection to mobile telephones. The well-known case of the bugged telephone conversations of the Prince of Wales comes to mind.

Today's widely popular radio-based computer networks (based on IEEE 802.11b), provide a four times higher risk factor than mobile phones:

- Data in computer networks cover a far greater bandwidth of content than telephone conversations.
- Because the data is stored electronically it is easily saved, filtered and searched.
- Networks have the potential to allow access to substantially more information. In many cases it is possible to bypass firewalls through Wave-LAN and thus gain access to a company's intranet.
- While expensive technology is required for listening in on mobile telephone conversations, with wireless networks an adaptation of the network software is sufficient.

In the case of an unauthorized access to a Wave-LAN, various scenarios have to be taken into consideration. Here are a few examples:

- **Internet Use:** An unauthorized person could use another persons' access to the Internet. This provides him with a high degree of anonymity. Besides sparing the cost of his – possibly punishable – acts, all suspicion will point to the person holding the access account.
- **Picking up Communication:** The unauthorized person may pick up communication within a wireless network and possibly even within the company's intranet.
- **"Data Theft":** Through the fact that the unauthorized person may access the data from "within the network" there is a great probability that he/she may read data stored within the network that may not be retrieved from outside.
- **Data Manipulation:** An unauthorized person may possibly change or modify

saved data or data that is being transmitted.

## 2 Technical Basics

Technology for today's prevailing wireless networks is the Wireless Ethernet in accordance to the IEEE standard 802.11b (Wave-LAN). These networks as *ad-hoc networks* enable several computers to communicate with each other equally, or in *managed mode*, in which a base station controls data transfer. We limit our report to the *managed mode* system, because it seems to be less complex and – as we will prove – more common.

Furthermore, we will briefly describe the technical functionality of Wave-LAN networks to form the basis for our legal assessment. Concurrently, we will outline the arbitrary mistakes published in technical literature.

### 2.1 Login

The base station of a Wave-LAN network sends several so-called *beacon* signals a second on an assigned channel. This *beacon* signal contains – besides other information on the network – the name (called *SSID*) and the worldwide unique address of the network and/or the base station. Because the addresses are assigned according to type, it is possible to find out the producer of the base station by this address. Furthermore, the *beacon* signal contains the information of whether encoding was used, of which channel was utilized, and of which speed it supports.

Control messages transmitted between the base station and a computer contain, amongst others: *probe request*, *probe response*, *authentication request* and *authentication response*. A computer will send a *probe request* to a specific *SSID* in order to receive information about the network. The base station of the relevant network will then answer with a *probe response*, which includes about the same data as the *beacon* signal. If no *SSID* is included in the *probe request* (also called *broadcast SSID*) all base stations will answer no matter which *SSID* they have.

A base station and thus the relevant network may however be configured as "hidden"; this is called *closed network access control*, *hidden SSID* or *no broadcast SSID*. If this set-up is chosen the base station will not include the *SSID* in their *beacon* signals and it will not answer to *probe request* messages to the *broadcast SSID*.

A try to limit access will fail however, because the name of the network will be transmitted in numerous control packets and may be easily detected by anybody who is able to receive data communicated between the base station and other computers.

Before a computer can transmit reference data to and from a base station it needs to authenticate itself at the base station with an *authentication request*. In the case of an *open system* configuration there is basically no authorization protocol. In a *shared key* authentication only those computers that own a password for that network, will be allowed access. However, an error in the protocol design will lead to the consequence that anyone, who will observe the successful authentication of another computer with the base station, will be able to simulate such an authentication without knowing the password. In this case, only two data packets that were recorded when another computer was successfully authenticated are mathematically linked in order to receive new valid authentication data [3].

Usually base stations offer another possibility to limit access and this is called *MAC Address Control*.[1] Every network interface card in computers has its own worldwide unique so-called *MAC* address. The base station administrator may exclude specific *MAC* addresses from access or it may limit access to specific *MAC* addresses. However, every user may change his or her computer's *MAC* address. With the help of specific software, he or she simply has to change the address of the network card to the value of an authorized card and thus be allowed access. To find out which addresses are authorized one either has to observe the communication or by automated trial and error try out all possible addresses of a producer.

### 2.2 Transport of Data

Maximum protection by encoding is intended for reference data. The Wave LAN Standard defines an encoding protocol called *WEP* (*Wired Equivalent Privacy*), which is intended to offer the same degree of safety to wireless networks that wired networks are standardized on.

---

[1] MAC: Medium Access Control.

Without going into technical detail, we would like name some of the possible ways of attack to the encoded transport of data:

- **Calculating the Password:** The utilized key can be calculated by recording the encoded data packet. This key opens the possibility to decode the entire recorded data as well as allowing the attacker to transmit data via the radio network [4].
- **Dictionary Attack:** By recording a substantial amount of encoded data packets, it is possible to decode single data packets without having to know the password [3].
- **Packet Modification:** Bits in data packets may be tipped. This allows the possibility – as long as the structure or content of the encoded data packet is known at least in part – to modify specific parts of the encoded data without knowing the code-key [3].
- **Packet Creation:** If the packets' encoded and unencrypted content is known – for example, in the case of an *authentication request* and *authentication reply* packet – then the ability to create an encoded package of any size is possible. The created package may be converted into any other encoded and valid packet [5].
- **Brute Force Attack:** Due to errors in the creation of code-keys, there is a good chance to just guess the password through trial and error. Only a few minutes of trying are sufficient for the implementations of some producers [6].
- **Replay Attack:** Attacks have been publicized that have made use of the fact that *WEP* encoding of an already encoded message will result in a decoded message. It possible to record an encoded message from the air and send it back to the radio network from outside via the base station. The base station will then try to re-encode it with the consequence that it then sends the unencrypted data [3].
- **Evil Twin:** A second base station with the same name but with greater transmitting power is installed. Most of the clients will now use the second base station. If the second base station is operated without encryption, the clients will often automatically deactivate encoding. Thus, the attacker will not only have access to unencrypted data, but also may falsify data passing his base station as he wishes – the so-called *man-in-the-middle attack* – and thus deactivate further security mechanisms, where required [7].

## 3 Methodology

In order to gain an insight on the spread and security configuration of Wave LANs and to enable an assessment of the expenditure to detect Wave LANs we have conducted a measuring campaign.

We chose the urban area of the city of Bonn as our testing area. In order to verify and extend the results gained from the Bonn area we have conducted sample measurements in the Cologne area.

In the preliminary test we used tradename notebooks, Wave LAN cards and exclusively the software supplied with it. For this preliminary test we chose the area around the faculty of law and political sciences in Bonn. In the network configuration of the two notebooks no network was specified. While we walked along the streets we researched, we had to check if our configuration software indicated a network. If so, we noted the location and the notebook was removed from the transmission range of the network. The network name was then reset to "unspecified".

In the course of this preliminary test we were able to register six Wave LAN networks in less than one hour.

To examine the complete urban area of Bonn, this method seemed inappropriate, as it was too slow and laborious. However, there is software on the market, which records network information and resets network names automatically. The most advanced software of this kind for the Microsoft Windows operating system is *Netstumbler,* which is available free of charge. For operating systems based on BSD 4.4lite *dstumbler* works in a similar way.

Each software carries out basically the same operations that were done manually in our preliminary test, and via an additional external GPS receiver they record the actual position. In addition they also record *hidden SSID* networks, which are usually sorted out by end user software.

The software will only passively detect the existence and the security configuration of a network by means of its control data but does not collect any encoded or unencrypted reference data.

In order to cover the entire urban area, the measurements were done by car. Our tests showed that it was possible to reliably detect networks at speeds of up to 65 km/h. Since this measurement set-up detected only two of the already known Wave LANs, we decided on a standard outside cars antenna with a performance of 7db – in order to counter the screening effect of the car body. With this antenna we were able to detect four of six known networks.

M. Dornseif and C. Klein carried out the following measuring campaign directly after the preliminary experiment from mid-October until the end of November 2001. In order not to disturb the traffic too much we regularly conducted our measurement tours in the late evenings or early mornings. That is why we could only detect networks that were also in operation overnight. While we do not assume that a significant share of base stations will be shut off over night we think that *ad hoc* networks consisting of computers only without any base station will presumably be switched off at night together with their constituent computers.

We tried to cover the central districts of Bonn. In the outlying districts we had to do with sampling measurement tours. All in all, we did about 20 measurement tours covering a total of almost 800 km.

## 4 Measurement Results

We could detect 157 networks in the urban Bonn area. Since we could only detect about 2/3 of the previously known networks with our equipment, we assume that the total number of networks (also the ones operated during usual business hours) in the Bonn area covered by the test run to about 200. When we made sample measurements in Cologne we detected another 125 networks, which were included in the summary. Only 11 (approx. 4 %) of these 283 networks were *ad hoc* networks; 78 networks (approx. 28 %) used *WEP* coding and 59 (approx. 21 %) were hidden (*hidden SSID*).

> This means that more than 50 % of the networks detected by us do not use the protocol security functions.

It is remarkable that 58 of the networks using *hidden SSID* (approx. 98 %) had the same network operator.

As we noticed 84 (approx. 30%) of the 283 networks we detected used the factory set names given by the various producers. Operators, who will not even change the name of their network, will presumably not adjust the security settings to their needs.

## 5 Evaluation with Respect to Criminal Law

A penal evaluation of the various methods of attacks to Wave LANs, which were briefed initially, is currently not being widely discussed. The main reason for that may be that attacks to radio-based computer networks would remain mostly undetected. Concurrently, there are probably a high number of undetected cases of computer crime – just like other white-collar crimes. Data espionage documented by the police for crime statistics, make up only 1% of all computer crimes in the year 2000. Indeed, in legal practice these attacks do not play a very important role.

### 5.1 § 202a StGB (German Criminal Code)

Wave LAN intruders may chose from a wide range of legally pertinent computer crimes. The relevant regulations of the German criminal code are § 202a and § 303a, b; whereas § 303a and b refer to § 202a with respect to the subject of the attack, which is the data. Thus, § 202a StGb represents the legal gate for attacks to radio networks.

This criminal provision has been introduced with the 2[nd] WiKG (Law on White-Collar Crime) of 1986 and was intended to close legal loopholes in respect to punishability, which have occurred at a time when the transfer of information has increasingly left the path of written communication; § 202 StGB, which has been created for cases of violations of privacy, and which exclusively applies to letters and secret documents, does not embrace today's information exchange taking place via computer networks. § 202a StGB – according to the prevailing yet not undisputed opinion – serves the protection of the personal spheres of life and secrecy as well as the general interest – formalized by the requirements for specific safety – in keeping data secret that perceivably cannot be stored or transferred directly.

According to the provision, the significant act here is that the offender obtains data for himself or for a third

Maximillian Dornseif, Kay H. Schumann, Christian Klein

person, which is not intended for him, and which was secured against unauthorized access.

The legislator has legally defined the subject of the crime – the data – in the second section of this provision. For the purpose of this provision: data that was stored or transferred electronically, magnetically or otherwise directly unperceivable. This applies to stored data and to that being transferred. In the case of an attack to a radio network with the purpose of actively or passively picking up the data transfer, the second alternative – data being transferred – applies. If the radio network is used only for the purpose of reading data stored on a single computer, the first alternative applies.

## 5.2 Spying out Stored Data

The criminally relevant objects of an attack when a radio network is invaded for the purpose of reading data stored on the computers of this network can be made out quickly and relevantly defined. Here too, however, is the limiting condition, which says that this data must be specially secured against unauthorized access. It depends on whether access barriers within the network counter the intrusion, or if there are access barriers protecting the data on the computers directly.

Such access barriers on the Wave LAN level may exist in *hidden SSID, shared key or MAC Address Control* systems. As mentioned above however, it is quite simple to bypass these access barriers with simple methods by a marginally experienced computer user.

This implies the question whether these safety systems actually represent an "exceptional security" with respect to the body of an offence as stipulated in § 202a StGB. In this regard there is a wide range of opinions to be found in pertinent literature.

Some would like to accept a special safety measure only if it is objectively suited to represent a considerable obstacle for very experienced computer experts [9, 10]. The majority here is of the opinion that it should be sufficient only if the experienced computer user or a non-professional expert will face a considerable obstacle. Although there has been no visible sign of a common pattern yet [11-18]. As we stated in our analysis of facts above, it is possible to access networks with very simple methods even for the non-professional. With respect to radio networks there are no sophisticated and at the same time practicable and reliable safety measures at present. Therefore, it should be sufficient, to use the interested non-professional as an example. This non-professional – when facing customary access barriers – must intensively grapple with the network, may have to calculate passwords, or modify its address data.

Thus, concerning these types of access barriers, the network operator should fulfill his obligation by providing adequate documentation in the interest of secrecy. As long as he has done this, he is lawfully protected against the retrieval of data from his computers.

## 5.3 Spying out of Data Being Transferred

With respect to attacks on data being transferred, in this case data that is subject to the actual radio transfer, the situation is similar. When discussing the issue whether the attack on the radio network represents a criminal act, one first has to examine whether the network operator had the data transmission secured physically, e.g. with screening measures. If this is the case, one can assume that special safety measures were made for the purpose of protection against a criminal act, and thus the data being transferred is lawfully protected.

If there are no special physical safety measures made, one must find out whether the operator has chosen the option of encoding (*WEP*). If this is not the case punishability according to § 202 StGB is fundamentally ruled out regarding the attacker because the transferred data had not been secured on a regular basis.

## 5.4 Legal Evaluation of Encoding

Looking at the wording of § 202a StGB however, the issue arises whether encoding can be considered a "special safety measure against unauthorized access", which represents an issue that has not been ultimately determined as yet in science and legislation.

The legislator indeed intended to protect data which are being transmitted by the creation of § 202a StGB. In this respect there should be no difference between the data transmission handled by cable or by radio.

However, if the operator uses data encryption, one must observe that the encoded data itself may be transmitted *unencrypted* without any further protection. The choice of words in § 202a StGB however, speaks of data that has been expressly secured against unauthorized access. However, the encoding itself merely creates new data, which must be transmitted. The original data, which is the subject of the encoding, continue to exist in their original form but are independent from the now self-contained transmission data being transferred. Possible attacks on the radio network will, however, aim at just these (encoded) data. Subject of this regulation in § 202a StGB are all data, which are subject to an interest to be kept secret. So with respect to the punishability, it does not make any difference whether a plain text message or an inextricable "mishmash" (of encoded data) is the target of the attack.

So we must take a look at the encoded data and not at the prototypes – the original data – standing somewhere in the background. Subsequently, if we now rightly comprehend the relevant transmission data as independent data, we will find it completely possible to just grab at some wholly unprotected data "out of the air".

In law literature regarding § 202a StGB, encoding is nevertheless regarded as a special safety measure [8, 15, 16, 20]. Some people say that having access to encoded data cannot be regarded as an access to the original data [16]. This is indeed correct, but is insignificant in our opinion due to the fact that the object of the attack is the specific encoded data alone.

Furthermore, in relevant literature there is dialogue about the sense and purpose of the standard. An access barrier for the purpose of the prevention of a criminal offence serves the purpose to protect the formal interest of secrecy. In the case of data transmission this may only be reached by encoding. Therefore, this law must be construed in a way that encoding as well must be understood as a means of security considering the crime [16]. Other authors reach the same conclusion with similar arguments. They all have one thing in common – is and when this issue is considered at all – that they mainly speak about the original data and that they don't assign the encoded data their own quality as the object of the criminal offence.

Only a few of them reject a special security for the purpose of § 202a StGB with respect to the transmission of encoded data [21]. In respect to the represented view here, that original data and encoded data being the same is not appropriate, we agree with their conclusion.

> So it shows that – when § 202a StGB is evaluated correctly and dogmatically – encoded data generally are not lawfully protected.

We will have to wait and see which direction legal practice in the future will go regarding attacks on radio networks. We fear that the courts will follow prevailing opinion and have the picking up of encoded data ruled by § 202a StGB. But this would be sooner supported by result-oriented considerations, because if you take a closer look at the wording of § 202a StGB you will find uncomfortable loopholes regarding the punishability.

# 6 Conclusion

We have proven that it is impossible to operate an attack-proof Wave LAN with the systems currently available on the mass market.

We have proven further that Wave LANs are considerably widespread with the majority seeming to be operated without any encoding or other means of security.

Finally, we have stated that operators of unprotected Wave LANs may in practice not expect any protection against the spying out of his data from the criminal courts and that they have only very little protection against active attacks to their networks.

Even in the case of networks using encoding, the situation with respect to the criminal law is dogmatic and highly unsatisfactory if their data is spied out. On the contrary, the law fundamentally protects against active attacks as long as the interest in secrecy is documented. This documentation function may be fulfilled by any of the Wave LAN security mechanisms.

Since there are sufficient publications stating that Wave LANs are unsafe, the user of a Wave LAN should reckon to be accused himself if his data fall into the wrong hands via a Wave LAN system.

> In conclusion, one must advise every Wave LAN user, who transmits data that is worth being protected, to seriously consider following the *Lawrence Livermore National Laboratory* and do without the use of Wave LAN use for reasons of security.